\documentclass[aps,prb,twocolumn,showpacs,superscriptaddress,floatfix]{revtex4-2}  
\usepackage{ucs}
\usepackage{natbib}
\usepackage{graphicx}  % needed for figures
\usepackage{bm}        % for math
\usepackage{amssymb}   % for math
\usepackage{amsmath}
\usepackage{color}
\usepackage{CJK}
\usepackage{times}
\usepackage{verbatim}
\usepackage{mathrsfs}
\usepackage{hyperref}
\usepackage{ulem}
\usepackage{physics}
\hypersetup{colorlinks = True, urlcolor=blue, linkcolor=blue, citecolor=blue}
\newcommand{\sig}[2]{\sigma^{#1}_{#2}}
\newcommand{\bg}{\begin{pmatrix}}
\newcommand{\ed}{\end{pmatrix}}
\newcommand{\dg}{\dagger}
\newcommand{\sg}{\sigma}
\newcommand{\al}{\alpha}
\newcommand{\bt}{\beta}
\newcommand{\kp}{\kappa}
\newcommand{\ld}{\lambda}
\newcommand{\gm}{\gamma}
\newcommand{\ep}{\epsilon}
\newcommand{\dt}{\delta}

\newcommand{\pair}[2]{\langle #1 \rangle_{#2}}

\usepackage{cancel}
\usepackage{comment}

\begin{document}

\title{Dynamics of vacancy-induced modes in the non-Abelian Kitaev spin liquid}
%\input author_list.tex       
% D0 authors (remove the first 3 lines
% of this file prior to submission, they
% contain a time stamp for the authorlist)
% (includes institutions and visitors)
\author{Wen-Han Kao}
%\email{kao00018@umn.edu}
\affiliation{School of Physics and Astronomy, University of Minnesota, Minneapolis, MN 55455, USA}

\author{G\'abor B. Hal\'asz}
\affiliation{Materials Science and Technology Division, Oak Ridge National Laboratory, Oak Ridge, TN 37831, USA} 
\affiliation{Quantum Science Center, Oak Ridge, TN 37831, USA}

\author{Natalia B. Perkins} 
%\email{nperkins@umn.edu}
\affiliation{School of Physics and Astronomy, University of Minnesota, Minneapolis, MN 55455, USA}

\date{\today}
\begin{abstract}
We study the dynamical response of vacancy-induced quasiparticle excitations in the site-diluted Kitaev spin liquid with a magnetic field. Due to the flux-binding effect and the emergence of dangling Majorana fermions around each spin vacancy, the low-energy physics is governed by a set of vacancy-induced quasi-zero-energy modes. These localized modes result in unique characteristics of the dynamical spin correlation functions, which intriguingly mimic the single-quasiparticle density of states and further exhibit a quasi-zero-frequency peak. By recognizing the potential observability of these local correlation functions via scanning tunneling microscopy (STM), we show how the STM response is sensitive to the local flux configuration, the magnetic field strength, and the vacancy concentration. Constructing a simple model of the localized modes, we also elucidate how the local correlation functions can be interpreted in terms of the hybridization between these modes.

\end{abstract}
\maketitle

\section{Introduction}
Quantum spin liquids (QSLs) are currently of large interest, in part
because of their non-Abelian quasiparticles that may enable topological quantum computing \cite{Kitaev2003,Kitaev2006,Savary2016,Knolle2017ARCMP,KnolleMoessner2019,
Motome2019JPSJ,Broholm2020}.
A famous example of  QSLs is  realized in the exactly solvable 
Kitaev spin model on the honeycomb lattice, which supports a gapped non-Abelian chiral
QSL in a magnetic field \cite{Kitaev2006}.
 In this non-Abelian Kitaev spin liquid (KSL) state, an isolated flux pins a topologically protected zero-energy Majorana mode, which is of great interest for quantum information processing. With site dilution, each isolated vacancy in the KSL binds a $Z_2$ flux and, consequently, is associated with a localized Majorana mode\cite{,Willans2010,Willans2011}. In fact, the introduction of vacancies into the non-Abelian KSL gives rise to an entire set of quasi-zero-energy Majorana modes from which the true zero mode emerges \cite{Kao2021vacancy,Kao2021localization,Vitor2022,takahashi2023nonlocal}. The localized nature of these vacancy-induced quasi-zero-energy modes simplifies their detection and manipulation, two key ingredients in proposed implementations of Majorana quantum computation.
Also, with the ability to deliberately introduce vacancies into candidate materials such as
$\alpha$-RuCl$_3$~\cite{Lampen2017,Do2018,Do2020,Imamura2023}, these systems have the potential to realize a scalable network of such vacancy-induced Majorana modes.

In the companion paper \cite{kao2023STM}, we put forth a proposal for the detection of the vacancy-induced quasi-zero-energy Majorana modes. We suggested using inelastic scanning tunneling microscopy (STM) setups in which a non-Abelian KSL, containing a finite density of spin vacancies, acts as a tunneling barrier between a tip and a substrate. Our results revealed  several important features in the STM response that originate from the quasi-zero-energy modes. Most notably, we observed a pronounced peak near zero bias voltage in the derivative of the tunneling conductance. Both the voltage position and the intensity of this peak exhibit an increase with the density of vacancies, indicating %the
fractionalized nature of the quasi-zero-energy modes. We also demonstrated that the STM response can effectively probe the single-particle density of states of the Majorana fermions in the non-Abelian KSL.
   
We further note that similar setups   were previously proposed to study the zero-energy Majorana modes in the clean non-Abelian KSL \cite{Konig2020,Udagawa2021,Knolle2020,Bauer2023} and in the non-Abelian KSL with an isolated pair of vacancies \cite{takahashi2023nonlocal}.
   For instance,
when two flux excitations are created and well separated from each other in the non-Abelian KSL, localized almost-zero-energy Majorana modes
are attached to each flux. 
The presence of these modes manifests as an additional, nearly zero-energy peak inside the bulk gap of the dynamical spin correlation function. 
As the 
support for
this peak comes only from on-site terms and nearest-neighbor
bonds directly enclosing the flux, the intensity of this peak in the STM response diminishes rapidly as the tip-to-flux distance increases \cite{Bauer2023}. 
It is also worth noting that  with the
recent strong experimental effort in imaging anyons with STM \cite{Papic2018},
these proposals for the non-Abelian KSL become more and more viable. Additionally, tunneling experiments recently performed on $\alpha$-RuCl$_3$ have yielded intriguing results concerning low-energy excitations  \cite{Yang2023}, which further underscores the feasibility and relevance of using STM for studying features of spin liquidity in real materials. Consequently, there is a need to gain a comprehensive understanding of the STM response in QSLs.

 In this paper, we delve into the intricate physics that underlies the STM response in the site-diluted non-Abelian phase of the Kitaev honeycomb model \cite{Kao2021vacancy}. Notably, this model remains exactly solvable even in the presence of vacancies.
  As demonstrated in prior research \cite{Knolle2020,Bauer2023,takahashi2023nonlocal,kao2023STM},
 the inelastic STM response, specifically  the
second derivative of the tunneling current with respect to the bias
voltage, is proportional to the dynamical spin
correlation function of the Kitaev spin liquid. 
 In the presence of vacancies, the STM setup allows us to investigate not only the bulk dynamical spin correlation function, which is probed when the tip is positioned away from vacancy sites, but also other kinds of dynamical spin correlation functions
 when the STM tip is positioned directly on top of a vacancy site.
  These  vacancy-related
 spin correlation functions probe either the usual ``bulk'' spin components or special ``dangling'' spin components that exhibit divergent local susceptibilities \cite{Willans2010,Willans2011}.

We study the behavior of both types of vacancy-related spin correlation functions, namely, those associated with bulk and dangling spin components. Exploring how these correlations evolve in response to variations in magnetic field strength, vacancy concentration, and the local flux environment, we reveal a pronounced difference in their behavior.
Bulk spin correlation functions, much like their counterparts in the clean Kitaev model  \cite{Baskaran2007,Knolle2014,Knolle2015},   are ultra-short-ranged, i.e., only on-site
 or nearest-neighbor correlation functions with the same spin component are nonzero. They exhibit a finite flux gap, a strong peak above
the flux gap, and a diffuse continuum at higher energies. 
Furthermore, when the time-reversal symmetry is broken by the magnetic field, the vacancy-related bulk spin correlation functions show a distinctive in-gap peak structure that reflects the vacancy-induced quasi-zero-energy modes.
In contrast, dangling spin correlation functions can be long-ranged, with no restrictions on the spin components or the distance. They also exhibit a set of low-energy peaks in the presence of a magnetic field, providing a complementary perspective on the quasi-zero-energy modes.

  Our study demonstrates that the distinct structures observed in the bulk and dangling dynamical correlations can be comprehended
 with the help of a simple tight-binding model describing hybridization
among various quasi-zero-energy  Majorana modes attached to each vacancy. 
Crucially, this model elucidates the emergence of the Majorana zero mode bound to each vacancy when the number of quasi-zero-energy modes is odd.
At the same time, it correctly explains the structures of the sharp peaks that are found inside the bulk gaps of both the bulk and the dangling spin correlation functions, and reveals how these peaks directly reflect the various vacancy-induced quasi-zero-energy modes.

\section{Site-diluted Kitaev honeycomb model}\label{sec:Kitaev}

\begin{figure}
\includegraphics[width=0.8\columnwidth]{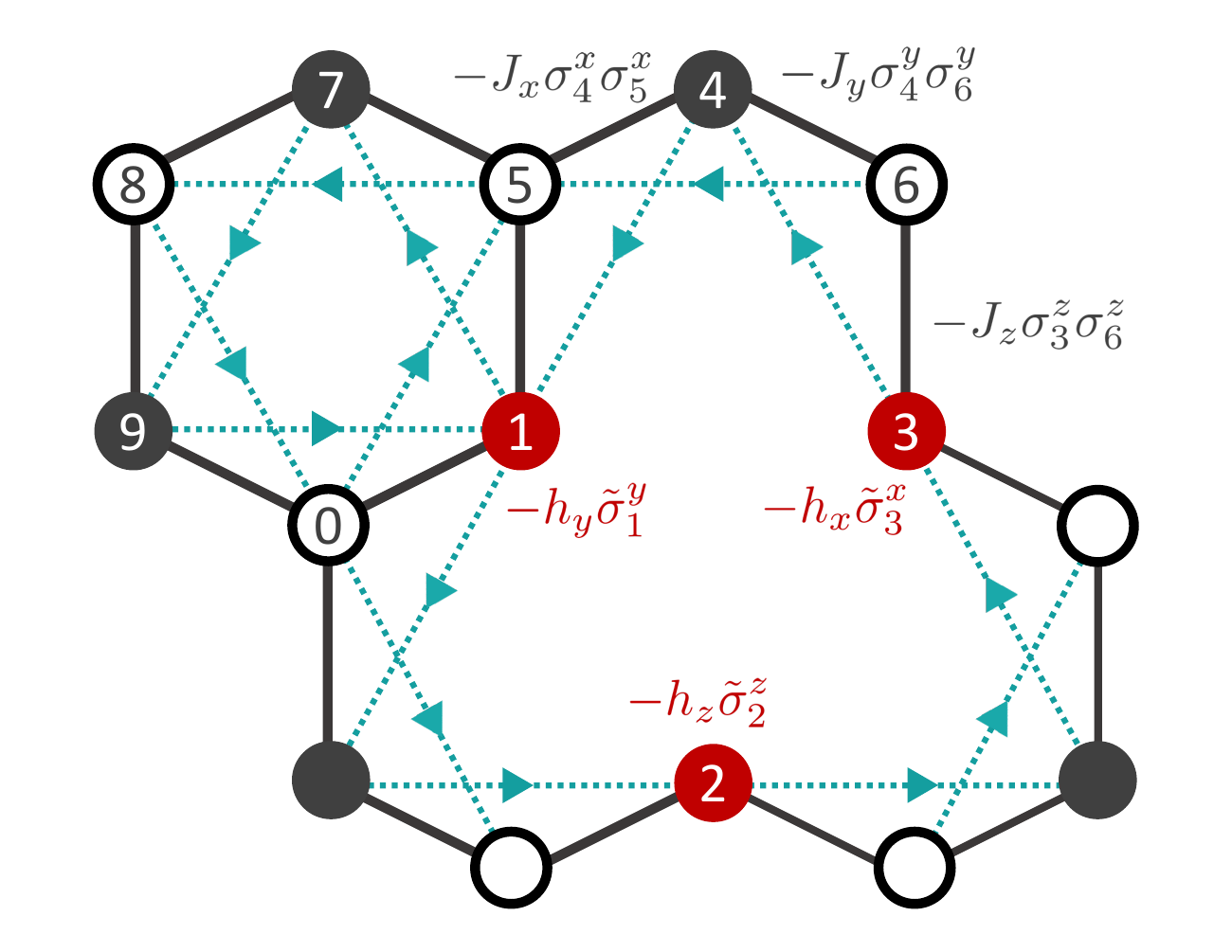}
     \caption{\label{fig:model} Site-diluted Kitaev honeycomb model. Around any vacancy, each of the sites 1, 2, and 3 has one dangling spin component $\tilde{\sigma}^{\alpha}_i$ that is directly coupled to the Zeeman field. The green dashed lines connect the sites coupled by the $\kappa$ term, for which the convention of chirality follows Ref.~\cite{Kitaev2006}. The hexagonal plaquette (flux) operator is defined as $W_p = \sig{z}{0}\tilde{\sg}_1^{y}\sig{x}{5}\sig{z}{7}\sig{y}{8}\sig{x}{9}$.}
\end{figure}

\subsection{Model Hamiltonian}
 We consider a site-diluted version of the Kitaev honeycomb model~\cite{Kitaev2006} where a finite number of spins are removed at random sites $j\in \mathbb{V}$, belonging equally to the A and B sublattices. The Kitaev interactions comprise nearest-neighbor bond-anisotropic Ising couplings emanating from each spin-1/2 degree of freedom. To account for a magnetic field, we also introduce the standard three-spin interactions~\cite{Kitaev2006} in the bulk, as well as the bare Zeeman terms $\propto \sg_j^{\al}$ at the sites $j \in \mathbb{D}_{\alpha}$ that share an $\alpha$ bond $\pair{jk}{\al}$ with a vacancy site $k\in \mathbb{V}$. By using Kitaev's four-Majorana representation, $\sig{\al}{j} = ib^{\al}_{j}c_{j}$, and the definition of $Z_2$ gauge fields, $u_{\pair{jk}{\al}} = ib^{\al}_{j}b^{\al}_{k}$, we obtain a quadratic fermionic Hamiltonian coupled to the gauge fields:
\begin{align}
\begin{split}\label{eq:Hamiltonian}
\mathcal{H} &= -\sum_{\pair{jk}{\al}}J_{\pair{jk}{\al}}\sg^{\al}_{j}\sg^{\al}_{k}-\kappa \sum_{\langle jkl\rangle_{\al\bt}}\sg^{\al}_{j}\sg^{\gm}_{k}\sg^{\bt}_{l}-\sum_{j \in \mathbb{D}_{\alpha}}h_{\alpha}\sg_{j}^{\alpha}\\
&= iJ\sum_{\pair{jk}{\al}}u_{\pair{jk}{\al}}c_j c_k+i\kappa\sum_{\langle jkl\rangle_{\al\bt}}u_{\pair{jk}{\al}}u_{\pair{kl}{\bt}}c_j c_l\\
& \quad\,\, -ih\sum_{j \in \mathbb{D}_{\al}}b^{\al}_{j}c_j,
\end{split}
\end{align}
where the summations are only over non-vacancy sites, and we assume $h_x = h_y = h_z \equiv h$ and
$J_{\pair{jk}{\al}} \equiv J$.
In terms of the Majorana fermions $c_j$, the $J$-term corresponds to first-neighbor hopping, whereas the $\kp$-term corresponds to second-neighbor hopping, as shown in Fig.~\ref{fig:model}. The flux operator $W_p$, which is a local symmetry defined on each plaquette, is related to the gauge fields by $W_p = \prod_{\langle jk \rangle_{\alpha} \in p} u_{\langle jk \rangle_{\alpha}} = \pm 1$.

%{\cred where  the first term corresponds to the pure Kitaev honeycomb model, the second term perturbatively incorporates a small magnetic field in the bulk of the spin liquid while preserving the exact solution of the pure Kitaev model~\cite{Kitaev2006}, and the  third term is the Zeeman  coupling included only on the on those sites  $\mathbb{D}_{\alpha}$ of the vacancy  on which dangling fermions reside, also preserving the exact solution of the model ~\cite{Kao2021vacancy, takahashi2023nonlocal}. The second and third terms explicitly  break time-reversal symmetry.}

From now on, we use a tilde to distinguish the ``dangling spin components'' $\tilde{\sg}^{\al}_j \equiv \sg^{\al}_j = i \tilde{b}^{\al}_{j} c_j$ with $j \in \mathbb{D}_{\alpha}$ and the associated ``dangling $\tilde{b}$-fermions'' $\tilde{b}^{\al}_{j} \equiv b^{\al}_{j}$ from the remaining spin components and $b$-fermions in the bulk. Since these dangling $\tilde{b}$-fermions do not have 
counterparts to form gauge fields,  they are treated as additional $c$-fermions in the tight-binding matrix of Eq.~(\ref{eq:Hamiltonian}).
In order to have the same structure of the tight-binding matrix as in the clean model, we assign each $\tilde{b}^{\al}_{j}$ to either the A or the B sublattice. We use the convention that the dangling $\tilde{b}$-fermions belong to the sublattice of the corresponding vacancy site. For example, if the vacancy is on the A sublattice, then  the neighboring dangling spin component $\tilde{\sg}^{\al}_{j}$ on the B sublattice  decomposes into $c_j$ on the B sublattice and $\tilde{b}^{\al}_{j}$ on the A sublattice. Therefore, the Zeeman term also looks like a nearest-neighbor hopping between A- and B-sublattice sites, and the Hamiltonian takes the general form
\begin{align}
\begin{split}
\mathcal{H} &= \frac{i}{2}
\bg c_A & c_B \ed \bg F & M \\ -M^T & -D \ed \bg c_A \\ c_B \ed\\
&= \frac{1}{2} \bg f^{\dg} & f \ed \bg \tilde{h} & \Delta \\ \Delta^{\dg} & -\tilde{h}^{T} \ed \bg f \\ f^{\dg} \ed,
\end{split}
\end{align}
where the complex matter fermions are defined as 
\begin{align}
\bg f \\ f^{\dg} \ed = \frac{1}{2}\bg 1 & i \\ 1 & -i \ed \bg c_A \\ c_B \ed.
\end{align}
Note that, in the site-diluted model, entries in the tight-binding matrix corresponding to vacancy sites are removed, and those corresponding to the dangling $\tilde{b}$-fermions are added. Therefore, the complex matter fermions are, in general, not formed by two Majorana fermions in the same unit cell. However, by the above assignment and the fact that only an equal number of vacancies on A- and B-sublattice sites are considered, each complex fermion still consists of one Majorana fermion from the A-sublattice and the other from the B-sublattice. The resulting Hamiltonian has the Bogoliubov de-Gennes form and thus can be diagonalized by a Bogoliubov transformation,
\begin{align}
\bg a \\ a^{\dg} \ed = \bg X^* & Y^* \\ Y & X \ed \bg f \\ f^{\dg} \ed, 
\end{align}
into the standard free-fermion form
\begin{align}
\quad \mathcal{H} = \sum_{n}\epsilon_n \left( a^{\dg}_n a_n - \frac{1}{2} \right),
\end{align}
with the excitation energies $\epsilon_n > 0$. The ground-state energy for a given flux sector is then $E_{0}(\{u\}) = -\frac{1}{2}\sum_n \epsilon_n$.

\subsection{Ground-state flux sector}
For the clean Kitaev honeycomb model, the Lieb theorem predicts that the ground-state flux sector is the \textit{zero-flux sector}, where all hexagonal plaquette operators have eigenvalue $+1$ \cite{Lieb1994}. In the general case, the ground-state flux eigenvalue (i.e., plaquette-operator eigenvalue) is determined by the number of sides $n$ for the plaquette:
\begin{align}
\langle W_p \rangle_{\mathrm{gs}} \sim e^{i\Phi_{\mathrm{gs}}}, \quad \Phi_{\mathrm{gs}} = \left(\frac{n-2}{2}\right)\pi \mod{2\pi}.
\end{align}
Interestingly, several works {\cite{Trebst2016,Willans2010,cassella2022exact} show that even without the requirement of reflection symmetry as used in the mathematical proof, numerically the ground-state flux eigenvalues still follow the prediction of the Lieb theorem. The most exotic case is the recently studied amorphous Kitaev spin liquid, in which the system contains all random shapes of plaquettes, and yet the ground-state eigenvalue of each plaquette is still determined by the Lieb theorem \cite{cassella2022exact}. 

For each isolated vacancy in the site-diluted honeycomb model, three hexagonal plaquettes merge into a single 12-site vacancy plaquette after removing one site at the center of the three hexagons. According to the Lieb theorem, the vacancy plaquette should carry a $\pi$-flux in the ground state, and this expectation has indeed been numerically verified in previous studies \cite{Willans2010, Willans2011, Kao2021vacancy, Vitor2022}. The corresponding ground-state flux configuration, where only vacancy plaquettes carry a $\pi$-flux, is called the \textit{bound-flux sector}. However, in the model of quasivacancies with the three-spin interaction $\kappa$, it was shown that a crossover from the bound-flux sector to the zero-flux sector is possible \cite{Kao2021vacancy}. Here, for true vacancies with dangling spins, we show that the ground-state flux sector exhibits a crossover even at $\kappa \sim 0$ as the strength of the Zeeman field $h$ on the dangling spins is increased [see Fig.~\ref{fig:FigureB2}(b)].

The algorithm for generating the bound-flux sector for random site-diluted configurations is based on our previous work \cite{Kao2021vacancy}. Additional care is taken to prevent a situation where two vacancy plaquettes are too close and merge into a larger vacancy plaquette because the evaluation of local spin correlations would be more complicated in such a case.

\begin{figure}
\includegraphics[width=1.0\columnwidth]{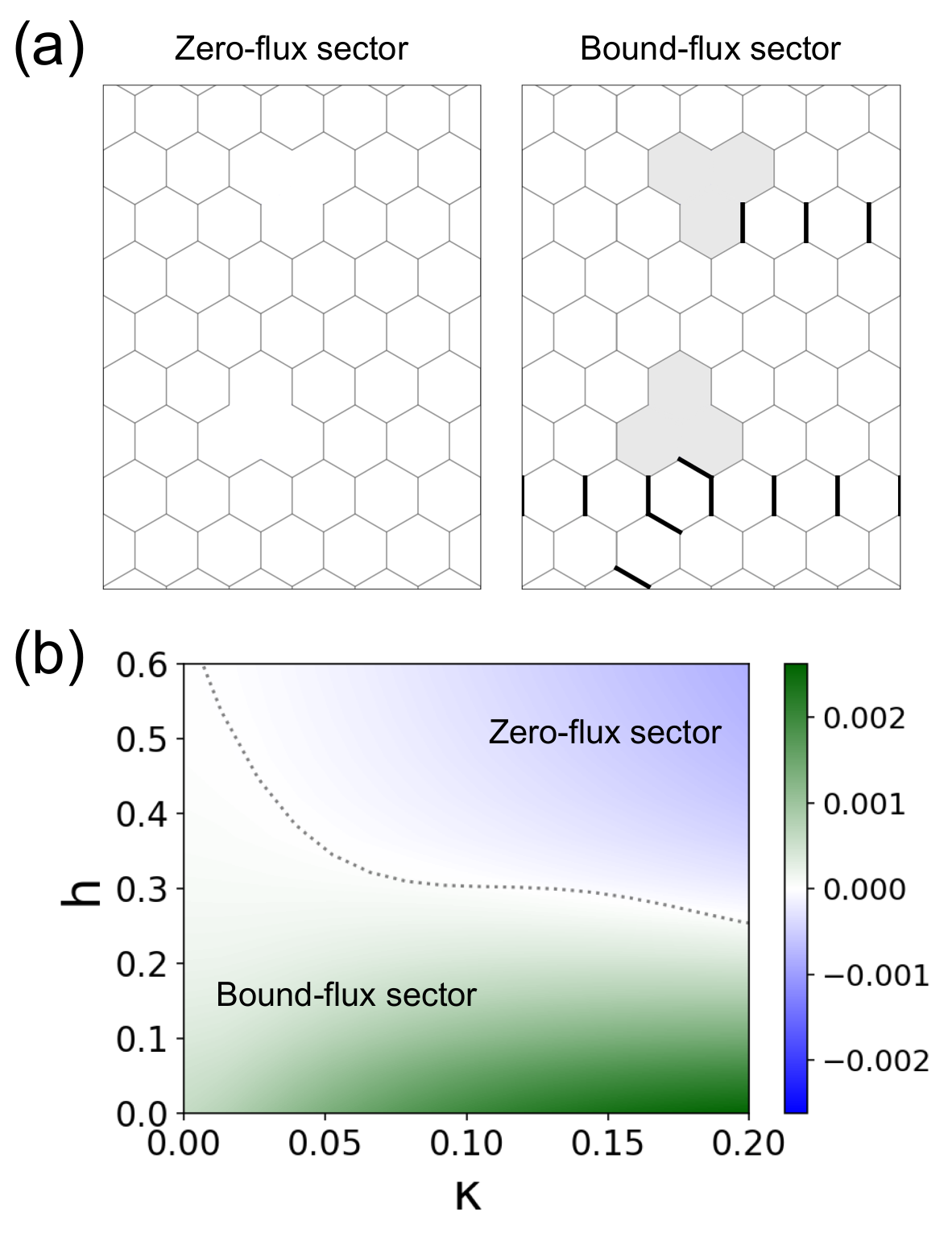}
     \caption{\label{fig:FigureB2} (a) Zero-flux and bound-flux sectors for the same vacancy configuration. In the bound-flux sector, the black thick lines mark flipped gauge variables $u_{\langle jk \rangle_\alpha}=-1$ and the shaded plaquettes indicate flipped flux operators $W_p = -1$. (b) Ground-state crossover between the zero-flux and the bound-flux sectors, as calculated from $L=40$ systems with $4\%$ vacancies. The shading corresponds to the energy difference between the two flux sectors.}
\end{figure}

\subsection{Local symmetry and expectation values}\label{sec:symmetry}
The presence of an extensive number of local symmetries in the Kitaev model can lead to predictions of vanishing expectation values of spin operators. The simple arguments presented below are related to the Elitzur theorem, which states that if an operator is not invariant under local symmetry transformation, then its expectation value must be zero \cite{Elitzur1975, Batista2005}.

First, we consider the clean Kitaev model without any vacancies. In this case, all minimum-size loops are hexagons, and each site (i.e., spin) belongs to three hexagonal plaquettes. From the definition of $W_p$  (shown in Fig.~\ref{fig:model}), we observe that any spin operator $\sg^{\al}_{j}$ anticommutes with at least one adjacent flux operator, $\{\sg^{\al}_{j},W_p\} = 0$, which means that it is not invariant under the corresponding local symmetry transformation: $W_p\sg^{\al}_{j}W_p = -\sg^{\al}_{j}W_p^2 = -\sg^{\al}_{j}$. Therefore, we conclude that the expectation value must be zero for all spins:
\begin{align}
\bra{0}\sg^{\al}_{j}\ket{0} = \bra{0}W_p\sg^{\al}_{j}W_p\ket{0} = -\bra{0}\sg^{\al}_{j}\ket{0} = 0.
\end{align}
Here $\ket{0}$ refers to the ground state of the Kitaev model in the zero-flux sector satisfying $W_p\ket{0}=\ket{0}$.

Similar arguments can be applied to the spin-spin correlation functions $\langle \sg^{\al}_j\sg^{\bt}_{l} \rangle$ \cite{Baskaran2007}. If the second spin $\sig{\bt}{l}$ is located beyond the nearest neighbor of $\sig{\al}{j}$, we can always find a flux operator that commutes with one spin and anticommutes with the other, $[\sg^{\al}_j,W_p] = \{\sg^{\bt}_l, W_p\} = 0$, which implies:
\begin{align}
W_p(\sg^{\al}_j\sg^{\bt}_{l})W_p = -\sg^{\al}_j\sg^{\bt}_l \quad \Longrightarrow \quad \bra{0}\sg^{\al}_j\sg^{\bt}_l\ket{0} = 0.
\end{align}
For a nearest-neighbor pair, if the spin components are not the same, we can still find a flux that anticommutes with only one of them, and hence $\bra{0}\sg^{\al}_j\sg^{\bt}_k\ket{0} = 0$ if $\al \neq \bt$. Therefore, the spin-spin correlation functions are ultra-short-ranged in the Kitaev model; only the on-site correlation functions and the nearest-neighbor correlation functions with the same spin component exhibit nonzero expectation values.
\begin{align}
\bra{0}\sg^{\al}_j\sg^{\al}_k\ket{0} \neq 0 \quad \mathrm{if} \,\,\, j = k \,\,\, \mathrm{or} \,\,\, j,k\in\expval{jk}_{\al}.
\end{align}

In the presence of vacancies, the above arguments for spin expectation values and spin-spin correlation functions remain valid, except for the dangling spin components $\tilde{\sg}^{\al}_j$ ($j\in\mathbb{D}_{\al}$) around each vacancy. The key point here is that no flux operator anticommutes with a dangling spin operator. Hence, any expectation values involving exclusively dangling spin components are generically non-zero:
\begin{align}\label{dangling-inequality}
&\bra{{\tilde 0}}\tilde{\sg}^{\al}_j\ket{{\tilde 0}} \neq 0, \quad \bra{\tilde 0}\tilde{\sg}^{\al}_j\tilde{\sg}^{\bt}_l\ket{\tilde 0} \neq 0.
\end{align}
Here $\ket{\tilde 0}$ refers to the ground state of the site-diluted Kitaev model, which may be in the bound-flux or the zero-flux sector. The non-zero expectation values in Eq.~(\ref{dangling-inequality}) provide the basis for the vacancy-induced local-moment formation \cite{Willans2010,Willans2011}
and show that spin-spin correlation functions between dangling spins have no restrictions in terms of spin components or distance. If $j$ and $l$ are around the same vacancy, we refer to $\langle \tilde{\sg}^{\al}_j\tilde{\sg}^{\bt}_{l} \rangle$ as an \textit{intra-vacancy} correlation function, otherwise we call it an \textit{inter-vacancy} correlation function. In this work, we focus on the intra-vacancy correlations, which can possibly be measured by a single STM tip \cite{kao2023STM}. The non-local behaviors of inter-vacancy correlations can be measured by a multiple-tip setup \cite{takahashi2023nonlocal}, but due to the localized wavefunctions of the vacancy-induced low-energy states, this inter-vacancy response is expected to be much smaller.

Finally, the spin-spin correlation functions between a bulk (i.e., non-dangling) spin and a dangling spin necessarily vanish because we can always find a flux operator that anticommutes with the bulk spin but commutes with the dangling spin, implying $\bra{\tilde 0}\sg^{\al}_j\tilde{\sg}^{\bt}_k\ket{\tilde 0} = 0$.

\section{Vacancy-induced localized modes}

\begin{figure*}
\includegraphics[width=1.0\textwidth]{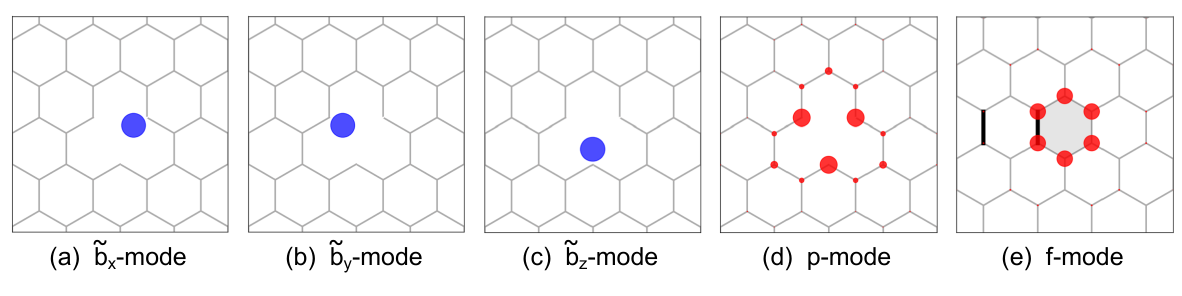}
     \caption{\label{fig:FigureB4} Different types of localized Majorana modes in the presence of vacancies and/or fluxes. (a-c) The vacancy-induced ${\tilde{b}}$-modes come from the fractionalization of the dangling spin components around vacancies. (d) The vacancy-induced peripheral mode (p-mode) has finite wavefunction amplitudes on the vertices of the vacancy plaquette. %only on the opposite sublattice with respect to the given vacancy. 
     (e) The flux-induced mode (f-mode) is localized around each $\pi$-flux plaquette in the non-Abelian Kitaev spin liquid.}
\end{figure*}

\subsection{Elementary modes and hybridization}

In the presence of the three-spin interaction $\kappa$ that mimics an external magnetic field and breaks time-reversal symmetry, the Majorana density of states becomes gapped, and the Majorana band structure acquires non-trivial topology. From the dispersion of the clean Kitaev model, the bulk Majorana gap
can be approximated by $\Delta_M \approx 6\sqrt{3}\kappa$ for $\kp \ll J$. In the resulting non-Abelian phase, if two fluxes are created and then separated from one another, a Majorana bound state with zero energy is supported and thus corresponds to an Ising anyon. These Majorana zero modes are highly localized around the $\pi$-flux plaquettes and are therefore called flux modes (f-modes) in this paper [see Fig.~\ref{fig:FigureB4}(e)].

When vacancies are introduced into the Kitaev spin liquid, two other types of localized modes emerge below the bulk gap. First, the unpaired $\tilde{b}$-Majorana fermions come from the fractionalization of dangling spins and couple to the bulk via the Zeeman term $h$. Each vacancy site exhibits three of these 
${\tilde{b}}$-modes, as shown in Fig.~\ref{fig:FigureB4}(a-c). Second, each vacancy gives rise to a peripheral mode (p-mode) around the vacancy position [see Fig.~\ref{fig:FigureB4}(d)]. This p-mode is similar to the quasilocalized mode with wavefunction amplitudes on the opposite sublattice sites to the vacancy, previously discussed in the tight-binding model of site-diluted graphene \cite{Pereira2006, Pereira2008}. However, in the KSL with $\kp \neq 0$, the wavefunction amplitudes of this p-mode are more localized and on both sublattices \cite{Kao2021vacancy}.

Since the wavefunctions of these $\tilde{b}$-, p-, and f-modes are all localized around the vacancy site, they can easily hybridize with each other, and the in-gap Majorana density of states is determined by the resulting eigenmodes [see Fig.~\ref{fig:FigureB5}]. In the bound-flux sector, the $\pi$-flux resides on the 12-site vacancy plaquette, and the f-mode wavefunction has a similar amplitude distribution to the p-mode wavefunction. Consequently, these two Majorana modes are strongly coupled by both the first-neighbor and the second-neighbor hopping terms of the Hamiltonian, and the energy of the resulting eigenmode [see Fig.~\ref{fig:FigureB5}(c)] can be written as $f(J,\kappa)$. These p-f hybridized modes correspond to the peak at $E \sim 1.16J$ in the density of states [see Fig.~\ref{fig:FigureB5}(a)]. In contrast, the peak at $E\sim 0$ is attributed to the $\tilde{b}$-modes that can only weakly couple to any other modes (including each other) via $h$. 

In the zero-flux sector, only three $\tilde{b}$-modes and one p-mode are available for each vacancy. Hence, two modes from the linear combinations of the $\tilde{b}$-modes are almost intact at $E\sim 0$ and the third one hybridizes with the p-mode to form a complex fermion of energy $E\sim h$ [see Fig.~\ref{fig:FigureB5}(f)]. The understanding of the in-gap spectrum based on the hybridization picture of localized Majorana modes is crucial for the later discussion on the dangling spin correlation function, so we discuss it in more detail in the next section.

\subsection{Simple model for hybridization}\label{toymodel}

\begin{figure*}
\includegraphics[width=1.0\textwidth]{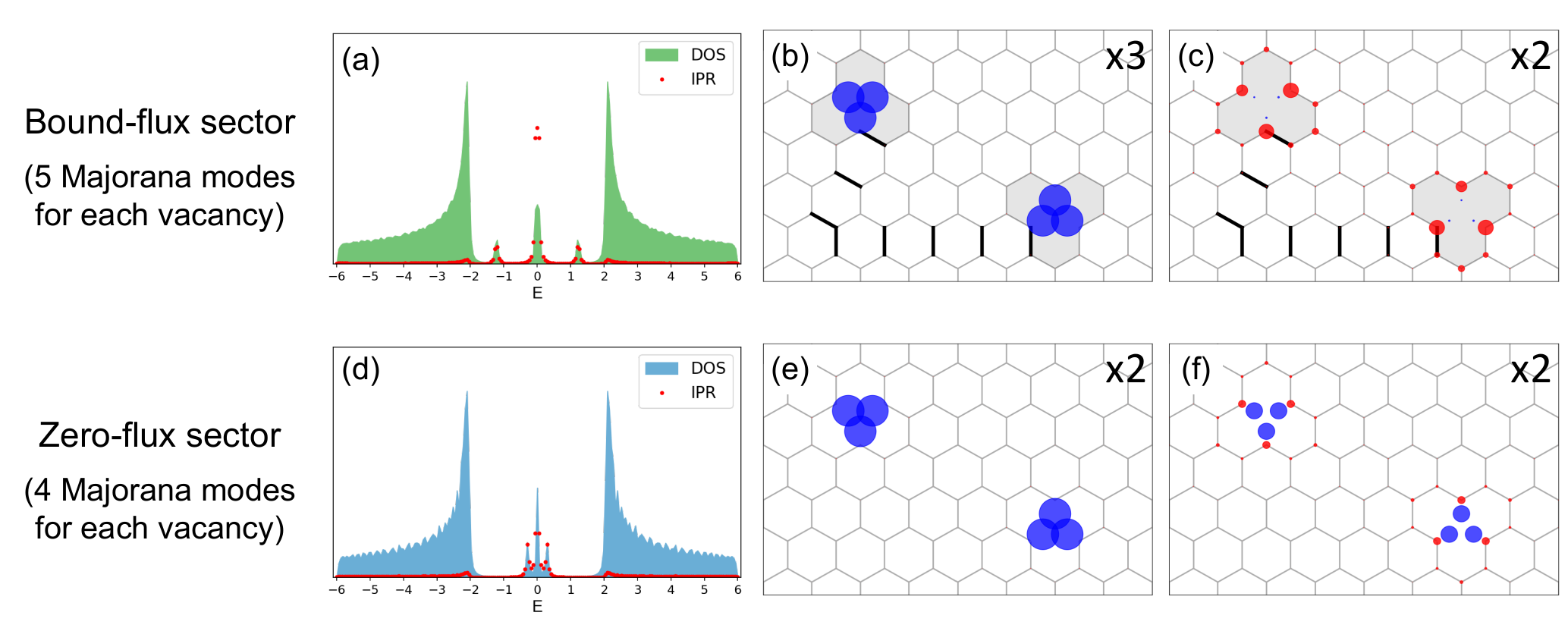}
     \caption{\label{fig:FigureB5} Density of states (DOS), inverse participation ratio (IPR), and absolute wavefunction amplitudes of in-gap eigenmodes in (a-c) the bound-flux sector and (d-f) the zero-flux sector. For the bound-flux sector, each vacancy contributes three $\tilde{b}$-modes to the central peak around $E = 0$ and two hybridized p-f modes to the finite-energy peaks around $E \approx \pm 1.16J$ in (a). For the zero-flux sector, each vacancy contributes three $\tilde{b}$-modes and one p-mode; one of the $\tilde{b}$-modes strongly hybridizes with the p-mode and thus a clear peak splitting is shown around $E = 0$ in (d). Numerical results in (a) and (d) are calculated from a $L = 40$ system containing 2\% vacancies with $h = \kappa = 0.2 J$. The intensity of IPR in (a) and (d) estimates the level of localization for the corresponding eigenmodes. It clearly shows that the vacancy-induced modes are much more localized than the modes in the continuum above the bulk gap.}
\end{figure*}

To develop a better understanding of the in-gap fermion modes, 
we analyze a simple tight-binding model describing the 
hybridization among the $\tilde{b}$-, p-, and f-modes attached to a single vacancy, and then compare the predictions of this effective model to the results from the exact diagonalization of the whole system with two well-separated vacancies.

We first consider the bound-flux sector, in which case the basis modes are $\tilde{b}^x,\, \tilde{b}^y,\, \tilde{b}^z,\, c_{\mathrm{p}},\, c_{\mathrm{f}}$. On the vacancy plaquette, the p-mode and the f-mode are strongly hybridized 
via both $J$ and $\kappa$ couplings, and the net hybridization can be described
by a homogeneous function $f(J, \kappa)$. The three $\tilde{b}$-modes, on the other hand, couple to the p- and f-modes by the Zeeman fields with hybridization strengths $\gm_\mathrm{p} h$ and $\gm_\mathrm{f} h$, respectively, where $\gm_{\mathrm{p}, \mathrm{f}} \sim 1$. Moreover, the $\tilde{b}$-modes can further hybridize among themselves via the higher-energy bulk modes (not included in the basis modes themselves). From second-order perturbation theory, we can estimate this hybridization as $\gm_b h^2$ with $\gm_b \sim J^{-1}$. Due to the threefold rotation symmetry relating the three $\tilde{b}$-modes, the effective Hamiltonian in the $(\tilde{b}^x,\, \tilde{b}^y,\, \tilde{b}^z,\, c_{\mathrm{p}},\, c_{\mathrm{f}})$ basis is then given by
\begin{align}
\mathcal{H}_{\mathrm{bound}} = i\bg 0 & \gm_b h^2 & -\gm_b h^2 & \gm_{\mathrm{p}} h & \gm_{\mathrm{f}} h \\ -\gm_b h^2 & 0 & \gm_b h^2 & \gm_{\mathrm{p}} h & \gm_{\mathrm{f}} h \\ \gm_b h^2 & -\gm_b h^2 & 0 & \gm_{\mathrm{p}} h & \gm_{\mathrm{f}} h \\ -\gm_{\mathrm{p}} h & -\gm_{\mathrm{p}} h & -\gm_{\mathrm{p}} h & 0 & f(J,\kappa) \\ -\gm_{\mathrm{f}} h & -\gm_{\mathrm{f}} h & -\gm_{\mathrm{f}} h & -f(J,\kappa) & 0  \ed.
\end{align}
Assuming $h \ll J$, the eigenvalues of $\mathcal{H}_{\mathrm{bound}}$ up to $O(h^2)$ are
\begin{align}
\epsilon \approx 0,\, \pm \sqrt{3}\gm_b h^2, \, \pm \left[
 f(J,\kappa)+\frac{3(\gm_{\mathrm{p}}^2+\gm_{\mathrm{f}}^2)}{2f(J,\kappa)
}h^2\right].
\end{align}
The first three eigenmodes have dominant $\tilde{b}$-fermion character, while the last two correspond to the p-f hybridized modes with mainly $c$-fermion character. 
The four finite-energy Majorana modes on each vacancy combine into two complex-fermion modes. However, the remaining zero-energy $\tilde{b}$-mode needs another zero-energy mode on a different vacancy to form a complex fermion. Therefore, this zero-energy mode is a protected Majorana bound state. To distinguish this true zero mode from the other hybridized $\tilde{b}$-mode with $h^2$ dependence, we call the former $\tilde{b}_0$-mode and the latter $\tilde{b}_h$-mode. In Fig.~\ref{fig:mode_hybridization}(a), we present the numerical eigenenergies of a full honeycomb-lattice system with two well-separated vacancies. The results confirm the picture from the simple model; the $\tilde{b}_0$-mode is clearly seen at the bottom of the spectrum, while the remaining two modes have quadratic $h$ dependence.
 
When a flux pair is excited from the bound-flux sector, as shown in Fig.~\ref{fig:mode_hybridization}(b), the p-mode and the f$^{\prime}$-mode are both zero modes at $h = 0$ because any hybridization between them is forbidden by symmetry~\footnote{Both modes transform the same way under reflection across the line that connects the vacancy site and the center of the f$^{\prime}$ plaquette, with the wavefunction amplitudes on the two sublattices having opposite parities. At the same time, the $J$-term connecting opposite sublattices is even while the $\kp$-term acting within individual sublattices is odd under the same reflection. Hence, the matrix elements of the Hamiltonian between the two modes necessarily vanish.}. Note that we use f$^{\prime}$ to denote the flux-induced mode that is localized on a normal plaquette instead of the vacancy plaquette. At $h > 0$, the $\tilde{b}$-modes couple to the p-mode on the vacancy plaquette and the f$^{\prime}$-mode on the normal plaquette. Given that only one $\tilde{b}$-mode can directly couple to the f$^{\prime}$-mode, the effective Hamiltonian in the $(\tilde{b}^x,\, \tilde{b}^y,\, \tilde{b}^z,\, c_{\mathrm{p}},\, c_{\mathrm{f}'})$ basis reads
 \begin{align}
\mathcal{H}^{\prime}_{\mathrm{bound}} = i\bg 0 & \gm_b h^2 & -\gm_b h^2 & \gm_{\mathrm{p}} h & \gm_{\mathrm{f}'} h \\ -\gm_b h^2 & 0 & \gm_b h^2 & \gm_{\mathrm{p}} h & 0 \\ \gm_b h^2 & -\gm_b h^2 & 0 & \gm_{\mathrm{p}} h & 0 \\ -\gm_{\mathrm{p}} h & -\gm_{\mathrm{p}} h & -\gm_{\mathrm{p}} h & 0 & 0 \\ -\gm_{\mathrm{f}'} h & 0 & 0 & 0 & 0  \ed,
\end{align}
and the corresponding eigenvalues expanded up to $O(h^2)$ are
\begin{align}
\epsilon \approx 0,\,\, \pm \gm_- h, \,\, \pm \gm_+ h
\end{align}
with $\gm_\pm^2 = \frac{1}{2} (3\gm_{\mathrm{p}}^2 + \gm_{\mathrm{f}'}^2 \pm \sqrt{9\gm_{\mathrm{p}}^4 - 2\gm_{\mathrm{p}}^2\gm_{\mathrm{f}'}^2 + \gm_{\mathrm{f}'}^4})$. The two complex-fermion modes with linear $h$ dependence can be interpreted as $\tilde{b}$-p and $\tilde{b}$-f$^{\prime}$ modes [see Fig.~\ref{fig:mode_hybridization}(b)], although in general each of them has both $\tilde{b}$-p and $\tilde{b}$-f$^{\prime}$ character. Again, the remaining $\tilde{b}_0$-mode is at exactly zero energy.

For the zero-flux sector, there are only four in-gap Majorana modes on each vacancy, and the effective Hamiltonian in the $(\tilde{b}^x,\tilde{b}^y,\tilde{b}^z, c_{\mathrm{p}})$ basis is given by
\begin{align}
\mathcal{H}_{\mathrm{zero}} = i\bg 0& \gm_b h^2& -\gm_b h^2& \gm_{\mathrm{p}} h \\ -\gm_b h^2& 0& \gm_b h^2& \gm_{\mathrm{p}} h\\ \gm_b h^2& -\gm_b h^2& 0& \gm_{\mathrm{p}} h\\ -\gm_{\mathrm{p}} h & -\gm_{\mathrm{p}} h & -\gm_{\mathrm{p}} h& 0\ed,
\end{align}
with the corresponding eigenvalues up to $O(h^2)$ being
\begin{align}
\epsilon \approx \pm \sqrt{3}\gm_b h^2,\,\, \pm\sqrt{3}\gm_{\mathrm{p}} h.
\end{align} 
Therefore, in the zero-flux sector, the simple model predicts a $\tilde{b}_h$-mode 
quadratic in $h$ and a $\tilde{b}$-p mode linear in $h$, but no protected zero mode. This prediction is confirmed by the numerical calculation on the whole system [see Fig.~\ref{fig:mode_hybridization}(c)]. 

Finally, when two fluxes are created from the zero-flux sector [see Fig.~\ref{fig:mode_hybridization}(d)], two additional in-gap f- and f$^{\prime}$-modes are introduced on two different plaquettes. Assuming that the hybridization at $h = 0$ is mainly between the p- and f-modes, the effective Hamiltonian at $h > 0$ in the $(\tilde{b}^x,\tilde{b}^y,\tilde{b}^z, c_{\mathrm{p}}, c_{\mathrm{f}}, c_{\mathrm{f}'})$ basis can be written as
\begin{align}
\mathcal{H}_{\mathrm{zero}}^{\prime} = i\bg 0 & \gm_b h^2 & -\gm_b h^2 & \gm_p h & \gm_f h &\gm_{f'}h \\ -\gm_b h^2 & 0 & \gm_b h^2 & \gm_p h & \gm_f h & 0 \\ \gm_b h^2 & -\gm_b h^2 & 0 & \gm_p h & \gm_f h & 0 \\ -\gm_p h & -\gm_p h & -\gm_p h & 0 & f(J,\kappa) & 0 \\ -\gm_f h & -\gm_f h & -\gm_f h & -f(J,\kappa) & 0 & 0 \\ -\gm_{f'}h & 0 & 0 & 0 & 0 & 0 \ed,
\end{align}
and the corresponding eigenvalues expanded up to $O(h^2)$ are
\begin{align}
\epsilon \approx \pm \gm_b h^2,\, \pm \gm_{\mathrm{f}'}h, \, \pm \left[
 f(J,\kappa)+\frac{3(\gm_{\mathrm{p}}^2+\gm_f^2)}{2f(J,\kappa)
}h^2\right].
\end{align}
The physical picture is that the p- and f- modes first form a complex fermion of energy $ f(J,\kappa)$ already at $h = 0$, then one of the $\tilde{b}$-modes couples to the f$^{\prime}$-mode with hybridization strength $\gm_{\mathrm{f}'} h$, and finally the remaining two $\tilde{b}$-modes hybridize via the second-order effect ($\gm_b h^2$) to give the $\tilde{b}_h$-mode with quadratic $h$ dependence [see Fig.~\ref{fig:mode_hybridization}(d)]. Note that considering an additional f-f$^{\prime}$ hybridization at $h = 0$ does not lead to a qualitative change of this picture.

In summary, the protected zero-energy mode only appears when the total number of vacancy-induced modes around each vacancy is odd, which is the case for the bound-flux sector. In contrast, for the zero-flux sector, the total number of vacancy-induced modes is even, and the lowest-energy eigenmode thus shows an $h^2$ dispersion instead of being at exactly zero energy. We note that these general statements are true not only for the ground states but also for the corresponding two-flux excited states that have the same parities of vacancy-induced modes.

\begin{figure}
\includegraphics[width=1.0\columnwidth]{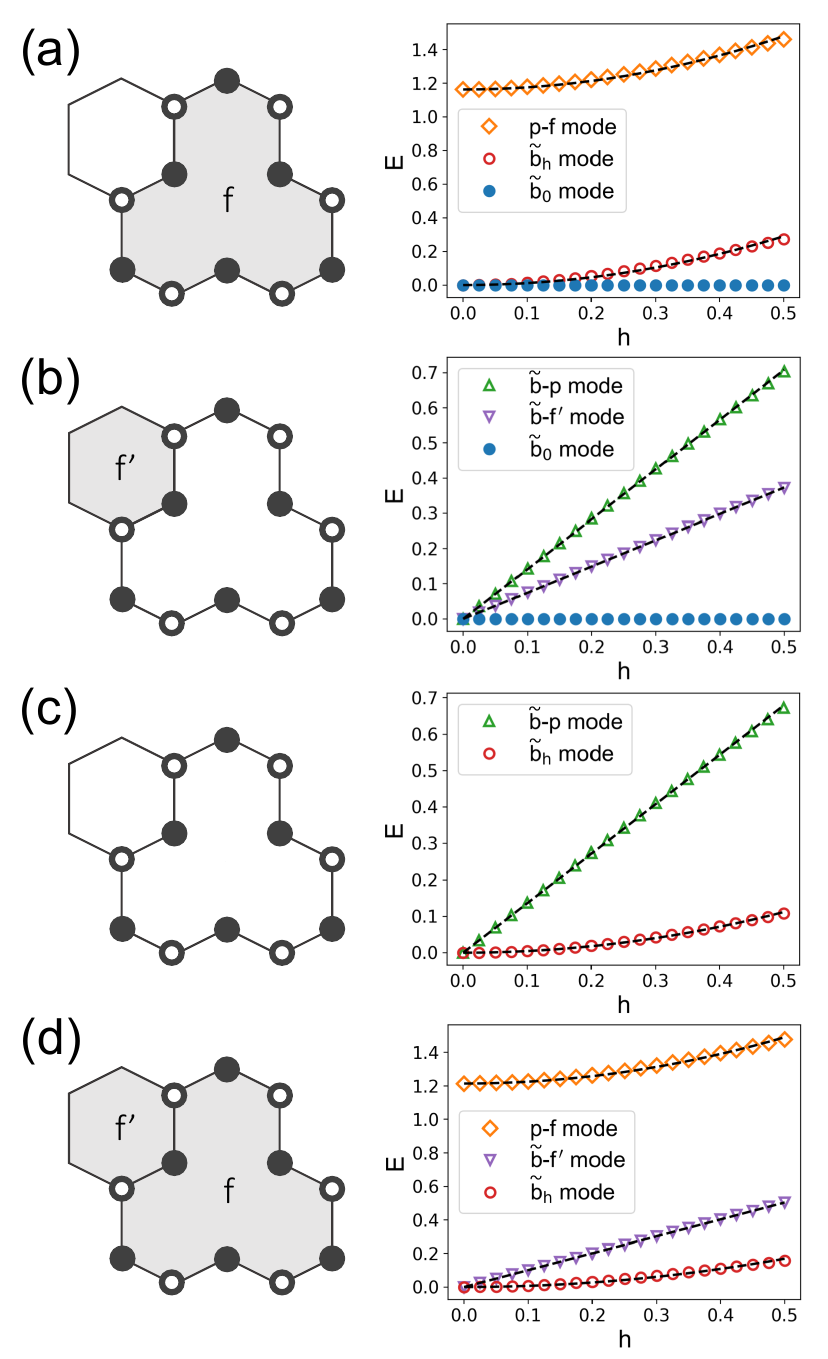}
     \caption{\label{fig:mode_hybridization} Different flux sectors around a vacancy and the corresponding fermion excitation energies. (a) Bound-flux sector. (b) Two flux excitations on top of the bound-flux sector; one flux is removed from the vacancy plaquette and another one is created on a neighboring hexagonal plaquette. (c) Zero-flux sector. (d) Two flux excitations on top of the zero-flux sector. For all subfigures, the f-mode (f$^{\prime}$-mode) denotes the flux-induced localized mode on the vacancy (hexagonal) plaquette.
     Excitation energies are calculated in $L = 40$ systems with two well-separated vacancies and $\kp = 0.2J$. The dashed lines are linear or quadratic fits in accordance with the simple model discussed in Sec.~\ref{toymodel}. The nomenclature of excitation modes in each subfigure is based on the dominant characters after hybridization.}
\end{figure}

\section{Dynamical correlation functions}\label{sec:correlator}

\subsection{Bulk spin correlation functions}
 We derive the dynamical spin correlation functions involving spin components in the bulk, and consider the ground state as a product state of the flux ground state and the fermion ground state, $\ket{0} = \ket{F}\otimes\ket{M}$. It is helpful to first define complex bond fermions in terms of the $b$-Majorana fermions:
\begin{align}
    \chi_{\pair{jk}{\al}} \equiv \frac{1}{2}(b_j^{\al}-ib^{\al}_k), \quad (j\in A,\, k\in B).
\end{align}
The number operator for this bond fermion is related to the $Z_2$ gauge-field operators by $u_{\pair{jk}{\al}} = 1-2\chi^{\dg}_{\pair{jk}{\al}}\chi_{\pair{jk}{\al}}$. Note that our definition is different from that in Ref.~\cite{Knolle2015}, because we 
%prefer
choose
the fixed gauge of the zero-flux sector ($u_{\pair{jk}{\al}} = 1$ for all bonds) to correspond to the vacuum of the bond fermions, that is  $\chi_{\pair{jk}{\al}}\ket{F} = 0$. This leads to the representation
of spin operators
\begin{align}
\sg^{\al}_j \equiv \eta_j \left(\chi^{\dg}_{\pair{jk}{\al}}+\zeta_j\chi_{\pair{jk}{\al}}\right)c_j,
\end{align}
where the sublattice-dependent prefactors are $(\eta_j, \zeta_j) = (i,1)$ for $j \in A$ and $(\eta_j, \zeta_j) = (1,-1)$ for $j \in B$.
The dynamical spin correlation function can then be derived as
\begin{align}
\begin{split}
S^{\al\bt}_{jk}(t) &= \bra{0}\sg^{\al}_j (t)\sg^{\bt}_k (0)\ket{0}\\
&= \eta_j\eta_k\zeta_j \bra{0}e^{i\mathcal{H} t}\chi_{\pair{jk'}{\al}}c_j e^{-i\mathcal{H} t}\chi^{\dg}_{\pair{j'k}{\bt}}c_k\ket{0}\\
&= \left(-\eta_j\eta_k\zeta_j\right)\bra{M}e^{i\mathcal{H} t}c_j e^{-i\mathcal{H}^{\prime}t}c_k\ket{M}\dt_{\al\bt}\dt_{\pair{jk}{\al}},
\end{split}
\end{align}
where we use the property
\begin{align}
\chi_{\pair{jk'}{\al}}e^{-i\mathcal{H}t} = e^{-i\mathcal{H}^{\prime}t}\chi_{\pair{jk'}{\al}}, \quad \mathcal{H}^{\prime} = \mathcal{H} -2iJc_jc_{k'}.
\end{align}
Note that we start with a general spin correlation function without assuming that $j$ and $k$ must be on the same bond. However, the selection rule of the flux sectors leads to the conclusion that the correlation function beyond nearest neighbors or with different spin components must be zero.

\subsubsection{One-particle contribution and adiabatic approximation}
For the nearest-neighbor correlation function with the same spin component, $\dt_{\al\bt}\dt_{\pair{jk}{\al}} = 1$, and by combining the sublattice prefactor as $\xi_{jk} \equiv -\eta_j\eta_k\zeta_j$, the correlation becomes
\begin{align}
S^{\al\al}_{jk}(t) = \xi_{jk}e^{iE_0 t}\bra{M}c_j e^{-i\mathcal{H}^{\prime}t}c_k\ket{M}.
\end{align}
Next, we apply the Lehmann representation for eigenstates of $\mathcal{H}^{\prime}$, and only keep the one-quasiparticle excitation sector $\{\ket{\ld} = (a^{\prime}_{\ld})^{\dg}\ket{M^{\prime}}\}$, where $\mathcal{H}^{\prime}\ket{\ld} = (E_0^{\prime}+\ep^{\prime}_{\ld})\ket{\ld}$ and $\ket{M^{\prime}}$ is the ground state of perturbed Hamiltonian $\mathcal{H}^{\prime}$. It was shown that the one-particle approach is a quantitatively good approximation since it contributes the most to the intensity under exact calculations \cite{Knolle2015}. We further apply the adiabatic approximation, and the correlation function in the frequency domain is then written as
\begin{align}
\begin{split}
S^{\al\al,(1)}_{jk}(\omega) = 2\pi\xi_{jk}\sum_{\ld}&\bra{M^{\prime}}c_j (a^{\prime}_{\ld})^{\dg}\ket{M^{\prime}}\bra{M^{\prime}}a^{\prime}_{\ld} c_k\ket{M^{\prime}}\\
&\,\,\times\dt[\omega-(E^{\prime}_0+\ep^{\prime}_{\ld}-E_0)],
\end{split}
\end{align}
 where the resonance frequency involves the two-flux excitation gap $\Delta_{2f} \equiv E^{\prime}_0 - E_0$.

To evaluate the matrix elements, we recall the transformation from $c$-Majorana fermions to complex fermions, and then to the Bogoliubov quasiparticles:
\begin{align}
c_j \rightarrow (-i\eta_j)(f_l+\zeta_j f_l^{\dg}), \quad f_l = \sum_{\ld}\left[X^{T}_{l\ld}a^{\prime}_{\ld}+Y^{\dg}_{l\ld}(a^{\prime}_{\ld})^{\dg}\right],
\end{align}
which leads to 
\begin{align}
\begin{split}
&\bra{M^{\prime}}c_j (a^{\prime}_{\ld})^{\dg}\ket{M^{\prime}} = -i\eta_j (X^{T}_{l\ld}+\zeta_j Y^{T}_{l\ld}),\\
&\bra{M^{\prime}}a^{\prime}_{\ld}c_k\ket{M^{\prime}} = -i\eta_k (Y^{\dg}_{m\ld}+\zeta_k X^{\dg}_{m\ld}).
\end{split}
\end{align}

By defining $W \equiv X+Y$ and $Z \equiv X-Y$, we  summarize the final results for the dynamical spin correlation function:
\begin{align}
\begin{split}
&S^{\al\al,(1)}_{j\in A,k\in A}(\omega) = 2\pi \sum_{\ld} W^T_{l\ld}W^{\dg}_{m\ld}\dt\left[\omega-(\Delta_{2f}+\ep^{\prime}_{\ld})\right],\\
&S^{\al\al,(1)}_{j\in A,k\in B}(\omega) = 2\pi \sum_{\ld} W^T_{l\ld}Z^{\dg}_{m\ld}\dt\left[\omega-(\Delta_{2f}+\ep^{\prime}_{\ld})\right],\\
&S^{\al\al,(1)}_{j\in B,k\in A}(\omega) = 2\pi \sum_{\ld} Z^T_{l\ld}W^{\dg}_{m\ld}\dt\left[\omega-(\Delta_{2f}+\ep^{\prime}_{\ld})\right],\\
&S^{\al\al,(1)}_{j\in B,k\in B}(\omega) = 2\pi \sum_{\ld} Z^T_{l\ld}Z^{\dg}_{m\ld}\dt\left[\omega-(\Delta_{2f}+\ep^{\prime}_{\ld})\right].\\
\end{split}
\end{align}

In the presence of vacancies, this derivation remains the same for the dynamical correlation function of the bulk spin components, and only the notation $\ket{M}$ is replaced by $|\tilde{M}\rangle$ as the fermion ground state of the site-diluted Kitaev model.

\subsection{Dangling spin correlation functions}
To compute the dynamical spin correlation functions for the dangling spins, $S^{\al\bt}_{jk}$, with $j \in \mathbb{D}_{\al}$ and $k \in \mathbb{D}_{\bt}$,
we treat the corresponding dangling $\tilde{b}$-Majorana fermions %are treated 
as $c$-Majorana fermions. They recombine into complex matter fermions instead of bond fermions as
\begin{align}
\tilde{b}_j^{\al} \rightarrow (-i\tilde{\eta}_j)(f_l+\tilde{\zeta}_j f^{\dg}_l).
\end{align}
It is important to note that the sublattice prefactor for dangling $\tilde{b}$-Majorana fermions has the opposite definition to the $c$-Majorana fermions, namely, $(\tilde{\eta}_j, \tilde{\zeta}_j) = (1,-1)$ if $j \in A$ and $(\tilde{\eta}_j, \tilde{\zeta}_j) = (i,1)$ if $j \in B$. This is because of our convention that, if $\tilde{\sg}_{j}^{\al}$ is on the A sublattice, the corresponding $c_j$ is defined on the A sublattice while $\tilde{b}_j^{\al}$ is defined on the B sublattice.

We can  then directly apply the Lehmann representation for the ground-state flux sector:
\begin{align}
\begin{split}
S^{\al\bt}_{jk}(t) &= \bra{\tilde{0}}\tilde{\sg}^{\al}_j(t)\tilde{\sg}^{\bt}_k(0)\ket{\tilde{0}}\\
&=-\sum_{\ld} e^{i(E_0-E_{\ld})t}\langle \tilde{M}|\tilde{b}^{\al}_j c_j|\ld\rangle\langle\ld|\tilde{b}^{\bt}_k c_k |\tilde{M}\rangle.
\end{split}
\end{align}
The leading terms are the zero-particle contributions with $|\ld\rangle = |\tilde{M}\rangle$ and $E_{\ld} = E_0$,
\begin{align}
S^{\alpha \beta, (0)}_{jk}(t) = -\langle \tilde{M}|\tilde{b}^{\al}_j c_j|\tilde{M}\rangle\langle \tilde{M}|\tilde{b}^{\bt}_k c_k|\tilde{M}\rangle,
\end{align}
and the two-particle contributions with $|\ld\rangle = a^{\dg}_{\gamma}a^{\dg}_{\dt}|\tilde{M}\rangle$ and $E_{\ld} = E_0+\ep_{\gm}+\ep_{\dt}$,
\begin{align}\label{eq:twoparticle-t}
\begin{split}
S^{\alpha \beta, (2)}_{jk}(t) = -\sum_{\gm, \dt}&e^{-i(\ep_{\gm}+\ep_{\dt})t}\langle \tilde{M}|\tilde{b}^{\al}_j c_j a_{\gm}^{\dg} a_{\dt}^{\dg}|\tilde{M}\rangle\\
&\quad\times\langle \tilde{M}|a_{\dt}a_{\gm}\tilde{b}^{\bt}_k c_k|\tilde{M}\rangle.
\end{split}
\end{align}

\subsubsection{Zero-particle contribution}
By transforming to the Bogoliubov quasiparticles, the %survived terms in the 
matrix elements 
%give
become
\begin{align}
\begin{split}
\langle \tilde{M}|\tilde{b}^{\al}_{j}c_j|\tilde{M}\rangle &= -\tilde{\eta}_j\eta_j\langle \tilde{M}|\left(f_l+\tilde{\zeta}_j f^{\dg}_l\right)\left(f_m+\zeta_j f_m^{\dg}\right)|\tilde{M}\rangle\\ &= -\tilde{\eta}_j\eta_j\sum_{\ld}\left(X^T_{l\ld}+\tilde{\zeta}_j Y^T_{l\ld}\right)\left(\zeta_j X^{\dg}_{m\ld}+Y^{\dg}_{m\ld}\right).
\end{split}
\end{align}
Therefore, the spin expectation values on the dangling sites belonging to the two sublattices are
\begin{align}
\begin{split}
&\langle \tilde{M}|\tilde{\sg}^{\al}_{j\in A}|\tilde{M}\rangle = \sum_{\ld}Z^T_{l\ld}W^{\dg}_{m\ld},\\
&\langle \tilde{M}|\tilde{\sg}^{\al}_{j\in B}|\tilde{M}\rangle = -\sum_{\ld}W^T_{l\ld}Z^{\dg}_{m\ld},
\end{split}
\end{align}
and the contributions to the correlation functions are appropriate products of these expectation values. We note that this zero-particle response only contributes to the elastic ($\omega = 0$) tunneling current in STM {\cite{kao2023STM}}.\\

\subsubsection{Two-particle contribution}
In the two-particle contribution, we need to evaluate four-fermion expectation values:
\begin{widetext}
\begin{align}
\langle \tilde{M}|\tilde{b}^{\al}_j c_j a^{\dg}_{\gamma}a^{\dg}_{\dt} |\tilde{M}\rangle = -\tilde{\eta}_j\eta_j \left[\left(X^T_{l\dt}+\tilde{\zeta}_j Y^T_{l\dt}\right)\left(X^T_{m\gamma}+\zeta_j Y^{T}_{m\gamma}\right)-\left(X^T_{l\gamma}+\tilde{\zeta}_j Y^{T}_{l\gamma}\right)\left(X^T_{m\dt}+\zeta_j Y^T_{m\dt}\right)\right].
\end{align}
By Fourier transformation, the dynamical spin correlation functions [see Eq.~(\ref{eq:twoparticle-t})] in the frequency domain become
\begin{align}
\begin{split}
&S^{\alpha\beta,(2)}_{j\in A,k\in A}(\omega) = 4\pi \sum_{\gm,\dt}\left( Z_{l\dt}^{T}W_{m\gm}^{T} \right) \left( Z_{l'\dt}^{\dg}W_{m'\gm}^{\dg}-Z_{l'\gm}^{\dg}W_{m'\dt}^{\dg} \right)\delta\left[ \omega-\left( \ep_\gm +\ep_\dt\right) \right],\\
&S^{\alpha\beta,(2)}_{j\in A,k\in B}(\omega) = 4\pi \sum_{\gm,\dt}\left( Z_{l\dt}^{T}W_{m\gm}^{T} \right) \left( W_{l'\dt}^{\dg}Z_{m'\gm}^{\dg}-W_{l'\gm}^{\dg}Z_{m'\dt}^{\dg} \right)\delta\left[ \omega-\left( \ep_\gm +\ep_\dt\right) \right],\\
&S^{\alpha\beta,(2)}_{j\in B,k\in A}(\omega) = 4\pi \sum_{\gm,\dt}\left( W_{l\dt}^{T}Z_{m\gm}^{T} \right) \left( Z_{l'\dt}^{\dg}W_{m'\gm}^{\dg}-Z_{l'\gm}^{\dg}W_{m'\dt}^{\dg} \right)\delta\left[ \omega-\left( \ep_\gm +\ep_\dt\right) \right],\\
&S^{\alpha\beta,(2)}_{j\in B,k\in B}(\omega) = 4\pi \sum_{\gm,\dt}\left( W_{l\dt}^{T}Z_{m\gm}^{T} \right) \left( W_{l'\dt}^{\dg}Z_{m'\gm}^{\dg}-W_{l'\gm}^{\dg}Z_{m'\dt}^{\dg} \right)\delta\left[ \omega-\left( \ep_\gm +\ep_\dt\right) \right].
\end{split}
\end{align}
%which 
The two-particle response provides the leading contribution to the inelastic tunneling current in STM {\cite{kao2023STM}}.
\end{widetext}

\subsection{Connected dynamical spin correlation function and disorder average}
Here we simplify the notations of the dynamical correlation functions for later usage and to be
consistent with the companion paper \cite{kao2023STM}. First, we define the connected dynamical spin correlation function as
\begin{align}
S^{\alpha\beta}_{c,jk} (\omega) = \int_{-\infty}^{+\infty} \frac{\mathrm{d}t}{2\pi} \, e^{i \omega t} \left[ \langle \sg^{\alpha}_j (t) \sg^{\beta}_k (0) \rangle - \langle \sg^{\alpha}_j \rangle \langle \sg^{\beta}_k \rangle \right].\label{eq:correlator}
\end{align}
For the bulk spin components, the second term vanishes due to $\expval{\sig{\al}{j}} = 0$ as discussed in Sec.~\ref{sec:symmetry}. For the dangling spin components, the second term simply cancels the zero-particle contribution, which means that $S^{\al\bt}_{c,jk}(\omega)$ corresponds to the two-particle contribution. Moreover, we restrict our attention to the on-site correlation $S^{\al\al}_{c,jj}(\omega)$ because it is more significant than the nearest-neighbor correlation. In the STM setup, the response from the nearest-neighbor correlation acquires an additional exponentially decaying prefactor with the distance between the two sites and is thus negligible. Finally, we define the disorder-averaged correlation functions
\begin{align}
\begin{split}
&\overline{S}_{\mathrm{bulk}}(\omega) = \overline{S^{\al\al,(1)}_{jj}}(\omega),\\
&\overline{S}_{\mathrm{dangling}}(\omega) = \overline{S^{\al\al,(2)}_{jj}}(\omega),
\end{split}
\end{align}
where the overline in $\overline{S}_{\mathrm{bulk}}(\omega)$ and $\overline{S}_{\mathrm{dangling}}(\omega)$ indicates averaging over different disorder realizations as well as over different vacancies in the same realization.

\section{Results and Discussion}

\subsection{Hybridization and dynamical spin correlation functions}

Here we describe the general connection between the dynamical spin correlation functions (see Fig.~\ref{fig:S_one_pair}) and the simple hybridization picture in Sec.~\ref{toymodel}. For the bulk correlation functions, the relevant matrix element is $\langle \tilde{M}^{\prime}|c_j (a^{\prime}_{\ld})^{\dg}|\tilde{M}^{\prime}\rangle$, which is calculated in the excited flux sector containing two flux excitations with respect to the bound-flux or zero-flux sector [see Figs.~\ref{fig:mode_hybridization}(b,d)]. Hence, there is a finite response at the frequencies $\Delta_{2f} + \ep^{\prime}_{\ld}$, where $\Delta_{2f}$ is the two-flux excitation gap and $\ep^{\prime}_{\ld}$ are the fermion excitation energies. However, since a finite response also requires that the fermion excitation $(a^{\prime}_{\ld})^{\dg}$ is compensated by a $c_j$ operator, the peaks corresponding to the $\tilde{b}_0$- and $\tilde{b}_h$-modes at small $h$ may not be discernible in the bulk correlation functions.

In contrast, for the dangling correlation functions, the relevant matrix element is $\langle\tilde{b}_j^{\al}c_j a^{\dg}_{\gm}a^{\dg}_{\dt}\rangle \equiv \langle\tilde{M}|\tilde{b}^{\al}_{j}c_j a^{\dg}_{\gm}a^{\dg}_{\dt}|\tilde{M}\rangle$, which is calculated in the ground-state flux sector:
\begin{align}\label{dangling}
S^{\al\al,(2)}_{jj}(\omega) = \sum_{\gm,\dt}\left|\langle\tilde{b}_j^{\al}c_j a^{\dg}_{\gm}a^{\dg}_{\dt}\rangle\right|^2 \dt[\omega-(\ep_{\gm}+\ep_{\dt})].
\end{align}
Therefore, we expect a finite response at all frequencies that correspond to sums of two distinct fermion excitation energies, $\ep_{\gm}+\ep_{\dt}$. Note that sums of two identical energies, $2\ep_{\gm}$, would correspond to $a^{\dg}_{\gm} = a^{\dg}_{\dt}$ and are thus prohibited by the Pauli exclusion principle. Also, a finite response requires that one of the excitations ($a^{\dg}_{\gm}$ or $a^{\dg}_{\dt}$) has finite $c$-fermion character and the other one has finite $\tilde{b}$-fermion character. 

As we discussed above, in
the {\it bound-flux} sector, there are three excitation modes on each vacancy: $\tilde{b}_0$ mode, $\tilde{b}_h$ mode, and p-f mode. Therefore, if the vacancies are well separated, we expect a finite response at three distinct frequencies:
\begin{align}\label{eq:omega123}
\begin{split}
&\omega_1 = \sqrt{3}\gm_b h^2,\\
&\omega_2 = f(J,\kp) +\frac{3(\gm_{\mathrm{p}}^2+\gm_f^2)}{2f(J,\kp)}h^2,\\
&\omega_3 =  f(J,\kp)+\left[\sqrt{3}\gm_b+\frac{3(\gm_{\mathrm{p}}^2+\gm_f^2)}{2f(J,\kp)}\right]h^2.
\end{split}
\end{align}
Note that, at very small $h$, both $\tilde{b}_0$- and $\tilde{b}_h$-modes have almost entirely $\tilde{b}$-fermion character and cannot compensate $c_j$ in the matrix element. Therefore, the peak at frequency $\omega_1$ is significantly weaker in intensity than the other two peaks.
As $h$ increases, however, the $\tilde{b}_h$-mode starts to hybridize more with $c$-fermions, and thus the $\omega_1$ peak gets stronger. Importantly, this peak appears at almost zero frequency and serves as direct evidence for the dangling $\tilde{b}$-fermions as well as the Majorana zero mode emerging from among them. 
 
Another route to enhance the peak at frequency $\omega_1$ is by introducing more vacancies to the system. With a higher density of vacancies and a shorter inter-vacancy distance, the dangling $\tilde{b}_0$- and $\tilde{b}_h$-modes can acquire stronger $c$-character and contribute more to the relevant matrix elements.

The presence of the true zero-energy mode (i.e., $\tilde{b}_0$-mode) in the bound-flux sector, engenders prominent features in the dangling correlation function of Eq.~(\ref{dangling}). If $a^{\dg}_{\gm}$ is the creation operator for the $\tilde{b}_0$-mode ($\ep_{\gm} = 0$) and has $O(1)$ overlap with a given $\tilde{b}^{\al}_j$ operator, the identity $(\tilde{b}^{\al}_j)^2 = 1$ gives
\begin{align}
S^{\al\al,(2)}_{jj}(\omega) \sim \sum_{\dt}\left| \langle\tilde{M}| c_j a^{\dg}_{\dt}|\tilde{M}\rangle \right|^2 \dt(\omega-\ep_{\dt}).
\end{align}
Thus, the dangling correlation function is largely proportional to the local density of states (LDOS) of the $c$-fermions, and it can directly capture the van Hove singularity at $E = 2J$ under weak fields. This behavior is distinct from the bulk correlation function because the latter involves a change in the flux sector and hence the response does not resemble the fermionic density of states \cite{Knolle2014}. The possibility of observing this van Hove singularity and the quasi-zero-frequency peak in the inelastic STM response is reported in the companion paper \cite{kao2023STM}.

In the case of the {\it zero-flux} sector, only two in-gap modes, $\tilde{b}_h$ mode and $\tilde{b}$-p mode, exist on each vacancy. As a result, we only expect a finite response at the frequency corresponding to the sum of their respective energies:
\begin{align}
\omega = \sqrt{3}\gm_{\mathrm{p}} h + \gm_b h^2.
\end{align}
Since the $\tilde{b}_h$ and $\tilde{b}$-p modes have significant $\tilde{b}$- and $c$-fermion characters, respectively, this response is pronounced even for small values of the Zeeman field $h$.

\subsection{Effect of local flux environment}

\begin{figure*}
\includegraphics[width=1.0\textwidth]{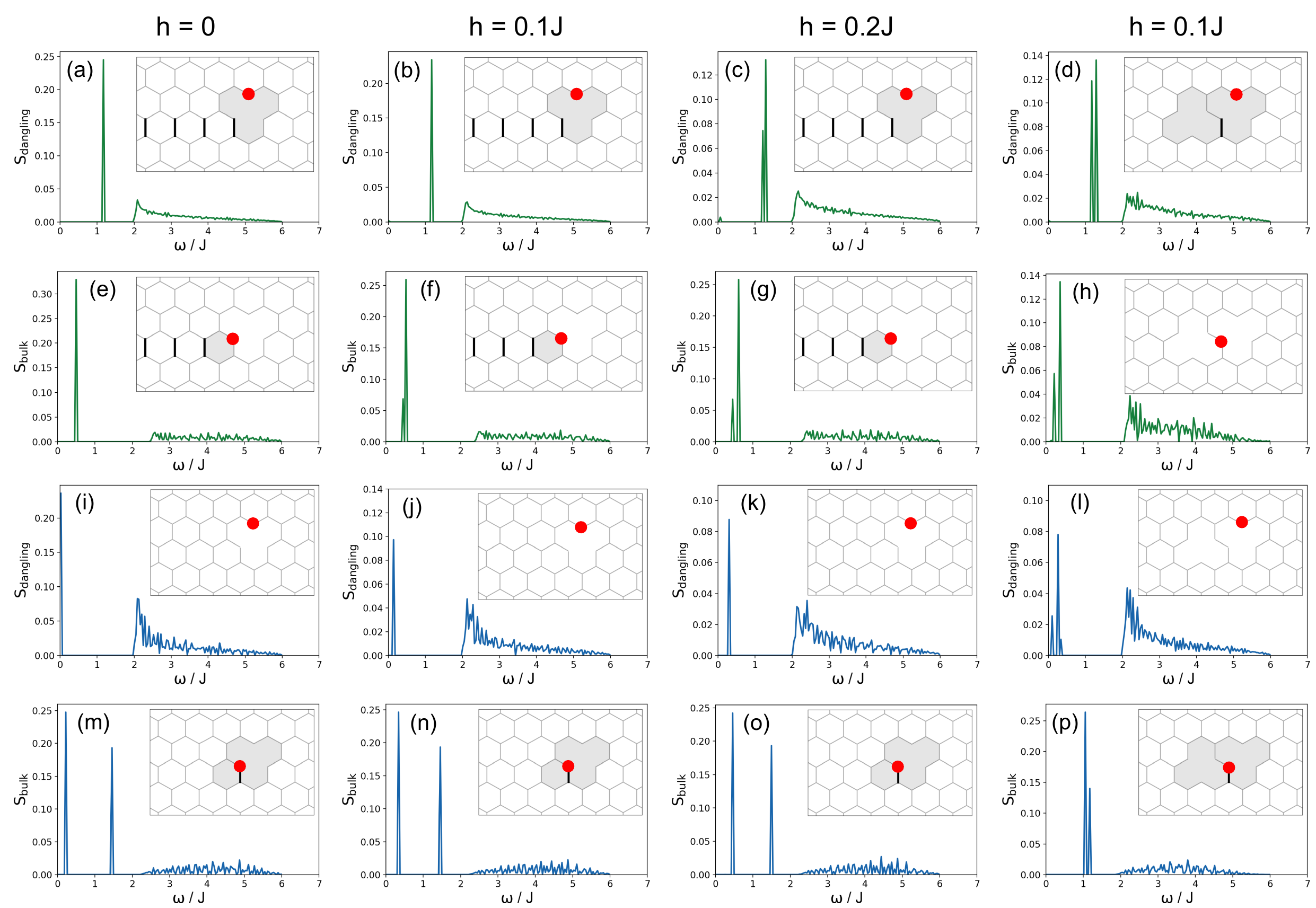}
     \caption{\label{fig:S_one_pair}Dynamical spin correlation functions around a vacancy in (a-h) the bound-flux sector and (i-p) the zero-flux sector. The parameter $\kappa = 0.2 J$ is chosen for all subfigures to separate the in-gap modes from the bulk response. The strength of the Zeeman field $h$ is shown at the top of each column. The inset of each plot shows the flux sector for calculating the matrix elements in the onsite correlation function $S^{zz}_{ii}(\omega)$ that corresponds to the lattice position $i$ denoted by a red dot.}
\end{figure*}

We first discuss the dynamical spin correlation functions in a system with only one pair of vacancies. In Fig.~\ref{fig:S_one_pair}, the parameter $\kappa$ is fixed at a relatively large value, $\kappa = 0.2J$, so that the bulk Majorana gap $\Delta_\mathrm{M} \approx 2J$ is well above the localized vacancy-induced modes. Therefore, all sharp responses at low energies are attributed to the hybridization among these localized modes.

For $S_{\mathrm{dangling}}$ in the bound-flux sector, only one sharp peak is observed for $h = 0$ at the frequency $\omega_2 = \omega_3 = f(J,\kp)$ [see Fig.~\ref{fig:S_one_pair}(a)], whereas the $\omega_1$ peak is absent due to the vanishing four-fermion expectation value $\langle\tilde{b}_j^{\al}c_j a^{\dg}_{\gm}a^{\dg}_{\dt}\rangle$. As $h$ increases, the quasi-zero-frequency peak at $\omega_1$ appears and, at the same time, the peak at $\omega_{2,3}$ splits, as already clearly seen in Fig.~\ref{fig:S_one_pair}(c). These two effects happen simultaneously because $\omega_3-\omega_2 = \omega_1 = \sqrt{3}\gm_b h^2$, i.e., both effects come from the hybridization of $\tilde{b}$- and $c$-fermions. In Fig.~\ref{fig:S_one_pair}(d), we consider the configuration where two vacancy plaquettes are in close proximity. In this case, the peak at $\omega_{2,3}$ shows stronger splitting due to flux-flux interaction \cite{Lahtinen_2011, Feng2020}. 

For $S_{\mathrm{bulk}}$ in the bound-flux sector, the creation of a flux pair means that the response is identically zero at frequencies below the two-flux excitation gap, $\omega<\Delta_{2f}$  \cite{Knolle2014,Knolle2015}. Note that this gap for annihilating the flux at the vacancy plaquette and creating another one at a neighboring hexagonal plaquette is $\Delta_{2f}\approx 0.48J$. For $h=0$, there is a single peak at the frequency $\omega = \Delta_{2f}$ [see Fig.~\ref{fig:S_one_pair}(e)], which implies that this peak originates from zero-energy modes: the p-mode on the vacancy plaquette and the f$^{\prime}$-mode on the neighboring hexagonal plaquette. For $h \neq 0$, these two zero modes both couple to the dangling $\tilde{b}$-modes, leading to a splitting of the sharp peak [see Fig.~\ref{fig:S_one_pair}(f-g)]. Based on our simple model of in-gap modes and the results from exact diagonalization of the whole system [see Fig.~\ref{fig:mode_hybridization}(b)], the energies of both the $\tilde{b}$-p mode and the $\tilde{b}$-f$^{\prime}$ mode have linear dependence in $h$, which means that this peak splitting also scales linearly with $h$.

For $S_{\mathrm{dangling}}$ in the zero-flux sector, both the $\tilde{b}$-modes and the p-mode are zero-energy modes at $h = 0$ but with different characters ($\tilde{b}$-fermion and $c$-fermion, respectively), which gives rise to a strong peak at $\omega = 0$ [see Fig.~\ref{fig:S_one_pair}(i)]. As $h$ increases, one $\tilde{b}$-mode hybridizes with the p-mode and the other two hybridize with each other through the bulk. Therefore, the peak in $S_{\mathrm{dangling}}$ corresponds to the creation of a $\tilde{b}_h$-mode as well as a $\tilde{b}$-p mode, and its frequency, $\omega = \sqrt{3}\gm_{\mathrm{p}} h + \gm_b h^2$, has dominant linear dependence in $h$ 
[see Figs.~\ref{fig:S_one_pair}(i-k)]. When the two vacancy plaquettes are in close proximity [see Fig.~\ref{fig:S_one_pair}(l)], the sharp peak splits into three peaks due to the hybridization of the $\tilde{b}$-p modes on the nearby vacancies.

For $S_{\mathrm{bulk}}$ in the zero-flux sector, there are two sharp peaks at well-separated frequencies, as shown in Fig.~\ref{fig:S_one_pair}(m-o). The lower-frequency peak corresponds to the $\tilde{b}$-f$^{\prime}$ mode whose energy is linearly proportional to $h$, while the higher-frequency peak corresponds to the p-f mode whose energy is already finite at $h=0$ [see Fig.~\ref{fig:mode_hybridization}(d)]. 
The frequencies of the two peaks are single-fermion energies with an additional shift from the two-flux gap, $\Delta_{2f} \sim 0.23J$.
Note that the response from the $\tilde{b}_h$ mode is almost invisible due to the vanishing two-fermion expectation value $\langle \tilde{M}^{\prime}|c_j (a^{\prime}_{\ld})^{\dg}|\tilde{M}^{\prime}\rangle$. 
Finally, we remark that when the two vacancy plaquettes are adjacent [see Fig.~\ref{fig:S_one_pair}(p)], the hybridization of the two modes
on the nearby vacancies with predominantly p-f character leads to two slightly split peaks appearing at the energy determined by $f(J,\kp)$.

The above analysis demonstrates how the dynamical correlation functions can reflect the local flux environment and detect the localized vacancy-induced modes. In systems with a small density of vacancies, bulk (e.g., spin-susceptibility) measurements would be dominated by the response from the non-disordered regions, thus being insensitive to the localized vacancy-induced modes \cite{Vitor2022}. However, real-space probes like scanning tunneling microscopy (STM) serve as a natural and informative tool for detecting these fractionalized modes and their low-energy density of states \cite{kao2023STM}.

\subsection{Effect of vacancy concentration}

\begin{figure*}
\includegraphics[width=1.0\textwidth]{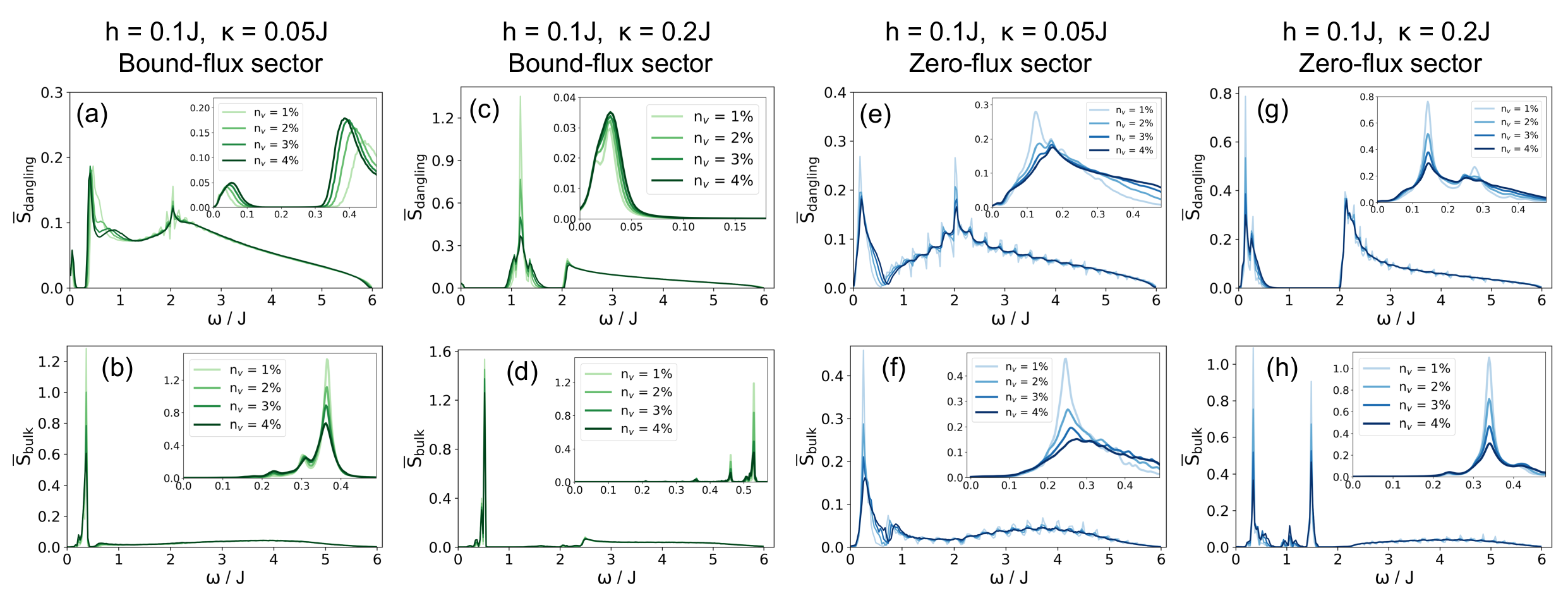} \caption{\label{fig:density_effect} Disorder-averaged dynamical spin correlation functions at various vacancy concentrations. All results are averaged over 500 disorder realizations with system size $L = 40$ in (a-d) the bound-flux sector and (e-h) the zero-flux sector. The insets of subfigures (a,c) emphasize the vacancy-concentration dependence of the quasi-zero-frequency peak.}
\end{figure*}

We now turn to the disorder-averaged correlation functions,    $\overline{S}_{\mathrm{bulk}}(\omega)$ and $\overline{S}_{\mathrm{dangling}}(\omega)$,  computed from multiple realizations for a finite concentration of vacancies. As in Ref.~\cite{kao2023STM}, we will only consider the dynamical spin correlation functions on the nearest-neighbor sites of the vacancies. These quantities determine the derivative of the tunneling conductance in STM when the tip is placed on top of the vacancy site. 

Figure~\ref{fig:density_effect} shows $\overline{S}_{\mathrm{bulk}}(\omega)$ and $\overline{S}_{\mathrm{dangling}}(\omega)$ as a function of the vacancy concentration in both the bound-flux and the zero-flux sectors, and for two different values of the three-spin interaction: $\kappa=0.05$ and  $\kappa=0.2$. For the larger value of $\kp$, different vacancies are decoupled from each other as the bulk correlation length is very small, and the results can thus be easily interpreted by the simple model for isolated vacancies discussed in Sec.~\ref{toymodel}. For the smaller value of $\kp$, however, the finite vacancy concentration is expected to be more physically relevant because a large number of vacancies are closer to each other than the bulk correlation length.

The most striking feature of a finite density of vacancies is the behavior of the quasi-zero-frequency peak in the bound-flux sector shown in Figs.~\ref{fig:density_effect}(a,c). We can clearly see from the inset of Fig.~\ref{fig:density_effect}(a) that, at small $\kp$, the frequency and the intensity of this peak both increase as the vacancy concentration grows. As discussed previously, this peak corresponds to frequency $\omega_1 = \sqrt{3}\gm_b h^2$, and its intensity is enhanced when the $\tilde{b}_h$-mode acquires more $c$-fermion character. Since a higher density of vacancies shortens the inter-vacancy distance and thus increases the hybridization, it leads to an intensification of the quasi-zero-frequency peak. In contrast, at large $\kp$, the wavefunctions of the $\tilde{b}_h$-modes are less hybridized with the bulk modes, and the dependence on vacancy concentration is thus less discernible [see Fig.~\ref{fig:density_effect}(c)]. Even though in practical experimental probes $\overline{S}_{\mathrm{dangling}}(\omega)$ and
$\overline{S}_{\mathrm{bulk}}(\omega)$ are measured simultaneously, the low-frequency signal in $\overline{S}_{\mathrm{bulk}}(\omega)$ does not hinder the visibility of the quasi-zero-frequency peak in 
$\overline{S}_{\mathrm{dangling}}(\omega)$
because of the flux gap.

The disorder-averaged results represent the typical behavior of random local configurations of vacancies, and thus more information can be extracted in specific cases. For example, in Fig.~\ref{fig:density_effect}(h), the disorder-averaged  $\overline{S}_{\mathrm{bulk}}(\omega)$ in the zero-flux sector shows two sharp peaks whose intensity decreases with the vacancy concentration.
These two sharp peaks have the same origin as in Fig.~\ref{fig:S_one_pair}(m), but they get broadened and less intense as the vacancy concentration increases because of interactions between localized modes on different vacancies. In addition, there is another broad peak with a much smaller intensity that appears between the two sharp peaks, whose origin can be understood in the same way as in
Fig.~\ref{fig:S_one_pair}(p). Its relatively smaller intensity reflects that adjacent vacancy plaquettes only rarely happen in the random local configurations. Once again, these results demonstrate the capability of the single-site dynamical spin correlation functions to probe the localized modes and the flux structure, which can then be potentially measured by an inelastic STM setup \cite{kao2023STM}.

\subsection{Effect of magnetic field}
In the site-diluted Kitaev honeycomb model [see Eq.~(\ref{eq:Hamiltonian})], we consider the Zeeman term applied only on the dangling spins
and the three-spin term applied only in the bulk, in order to preserve the exact solvability of the model. Practically, both terms come from the magnetic field, but in the presence of vacancies the relationship between them is rather obscure and depends on the microscopic details \cite{Masahiko2020}. Nevertheless, to see the quasi-zero-frequency peak, the simple guidance is that both terms should be present and be much smaller than the scale of $J$. In Fig.~\ref{fig:field_effect}, we also show that these two terms have opposite effects on the quasi-zero-frequency peak.

The $h$-dependence of $\overline{S}_{\mathrm{dangling}}$  with fixed $\kp = 0.05J$ is shown in Fig.~\ref{fig:field_effect}(a). At $h = 0$, the quasi-zero-frequency peak is absent due to the lack of hybridization between $c$ fermions and $\tilde{b}$ fermions. As $h$ increases, the peak is not only moved to higher frequencies but is also significantly broadened, which also indicates stronger hybridization at larger $h$. We also note that the small $\kappa$ used in Fig.~\ref{fig:field_effect}(a) makes the second peak overlap with the bulk continuum due to strong hybridization with the bulk $c$-fermions. This peak, originating from $\omega_2$ and $\omega_3$ in Eq.~(\ref{eq:omega123}), simply merges into the continuum at larger $h$.

In contrast, the $\kp$-dependence of $\overline{S}_{\mathrm{dangling}}$ with $h = 0.1J$, shown in Fig.~\ref{fig:field_effect}(b), clearly demonstrates that a larger $\kp$ (i.e., larger gap) corresponds to smaller hybridization, which means that the quasi-zero-frequency peak moves to lower frequencies and becomes sharper as $\kp$ is increased.  The second peak, which appears at the edge of the continuum, shifts to higher frequencies with increasing $\kp$ because $\Delta_M \approx 6\sqrt{3}\kp$. However, for $\kp = 0.06J$, we already see that the true edge of the continuum moves faster and starts to separate from the second peak. For large enough $\kp$, this peak is completely isolated from the bulk continuum, as shown in Fig.~\ref{fig:density_effect}(c). 

To summarize, we show that the quasi-zero-frequency peak reveals very distinct dependence on $h$ and $\kp$, because the hybridization involving dangling fermions is enhanced by the former and impeded by the latter. In the experiments, both the $h$ and the $\kp$ terms originate from the applied magnetic field, and thus the field dependence of this peak measured in STM can also help us to elucidate the relationship between the two effective terms in the Hamiltonian.

\begin{figure}
\includegraphics[width=1.0\columnwidth]{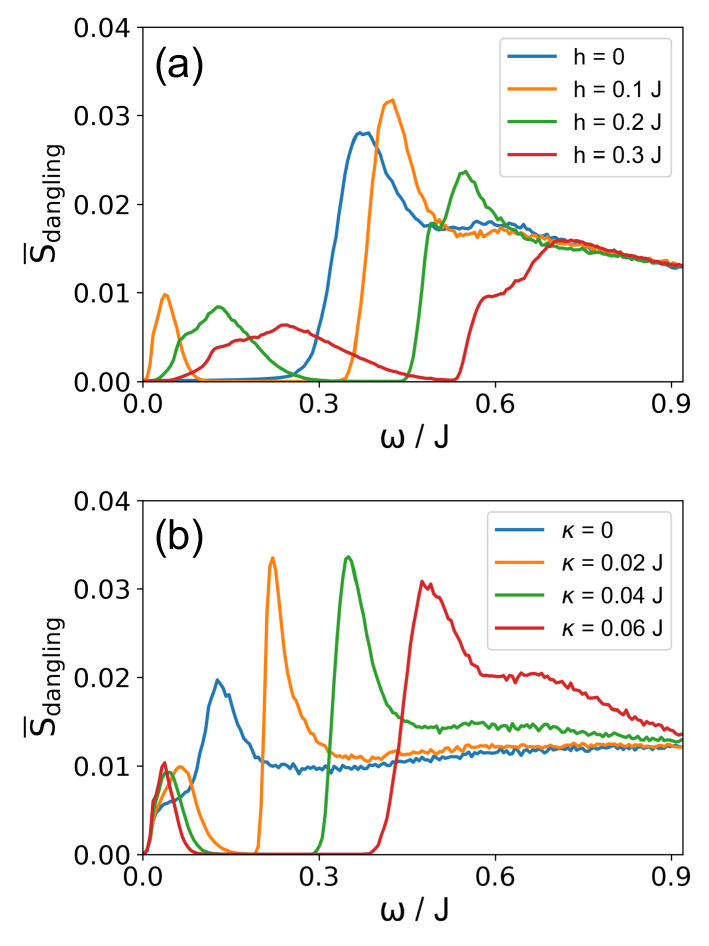}
     \caption{\label{fig:field_effect} The (a) $h$-dependence and (b) $\kp$-dependence of the dangling spin correlation function in the bound-flux sector. The strength of $\kp$ is fixed at $0.05J$ in (a) and the strength of $h$ is fixed at $0.1J$ in (b). The vacancy concentration is 2\% for both plots. All results are averaged over 500 disorder realizations with system size $L = 40$.}
\end{figure}

\section{Conclusion}
We study the dynamical spin correlation functions of the site-diluted Kitaev honeycomb model with the three-spin interaction $\kappa$ in the bulk and the Zeeman field $h$ coupled to dangling spins next to vacancies.
In the clean model, the three-spin interactions, which imitate the leading-order effect of the magnetic field, open a gap in the Majorana spectrum and engender non-trivial band topology. If two flux excitations are created and then well separated from each other, a localized Majorana mode is attached to each $\pi$-flux and behaves as a Majorana bound state inside the bulk gap \cite{Kitaev2006,Lahtinen_2011}. In the presence of vacancies, other types of in-gap modes can be induced around the vacancy positions. For example, the three dangling $\tilde{b}$-Majorana fermions, which originate from the fractionalization of dangling spins near the vacancies, can couple to the $c$-Majorana fermions through the Zeeman field. If the total number of localized Majorana modes around a given vacancy is odd, and these localized modes are well separated from the bulk modes due to the gap, one of these Majorana modes has to form a complex fermion with a Majorana mode on a different vacancy, and this mode is thus a protected zero-energy Majorana mode. This odd-number scenario occurs in the bound-flux sector, which is the ground-state flux configuration of the site-diluted Kitaev model. Therefore, in the presence of vacancies, protected zero modes can exist in the ground-state flux sector instead of excited flux sectors.

A natural question is then: how can we detect these zero modes induced by vacancies?  We attempted to answer this question in the companion paper \cite{kao2023STM}, where we have shown that the derivative of the tunneling conductance in inelastic STM is proportional to the dynamical spin correlation function around the tip position and can thus serve as vacancy spectroscopy. In this paper, we show that, next to a vacancy, the dynamical response can be well explained by our simple model of the in-gap modes. We also demonstrate that the local dynamical response is profoundly affected by the local flux environment, the vacancy concentration, and the magnetic field through the hybridization between the localized modes. Most importantly, the presence of the protected zero-energy mode makes the dangling spin correlation function resemble the local Majorana density of states and also gives rise to a quasi-zero-frequency peak. This peak can be potentially detected by STM because it is located in the no-intensity region of the bulk response. Considering the growing interest in disordered quantum spin liquids and their candidate materials with imperfections, our approach based on the local dynamical response around defects may be generalized to a wider class of spin-liquid systems for detecting defect-induced localized excitations and spin fractionalization.

\section{Acknowledgments}
 We would like  to thank
 Jason Alicea, Johannes Knolle, Patrick Lee, An-Ping Li, Abhay  Pasupathy  and Alan Tennant for enlightening discussions.
W.-H. Kao and N. B. Perkins acknowledge the support from NSF DMR-2310318 and  the support of the Minnesota Supercomputing Institute (MSI) at the University of Minnesota.
G.~B.~H. was supported by the U.S. Department of Energy, Office of Science, National Quantum Information Science Research Centers, Quantum Science Center. This research was supported in part by the National Science Foundation under Grants No.~NSF PHY-1748958 and PHY-2309135.

%\bibliographystyle{apsrev4-1}
%\bibliography{snakeref} 
\bibliography{STM_refs.bib}
\end{document}